\documentclass[12pt]{article}
\begin{document}
\title{On Vlasov approach to tokamaks near magnetic axis}
\author{H. Tasso \\ Max-Planck-Institut f\"{u}r Plasmaphysik,
Euratom Association, \\D-85748 Garching, Germany\\ \\
G.N. Throumoulopoulos\\
University of Ioannina, Association Euratom - Hellenic Republic,\\
Section of Theoretical Physics, GR 451 10 Ioannina, Greece}

\maketitle
\begin{abstract} A previous proof of non existence of tokamak equilibria with
purely poloidal flow within macroscopic theory [Throumoulopoulos,
Weitzner, Tasso, Physics of Plasmas {\bf 13}, 122501 (2006)]
motivated this microscopic analysis near magnetic axis for toroidal
and "straight" tokamak plasmas. Despite the new exact solutions of
Vlasov's equation found here, the structure of macroscopic flows
remains elusive.
\end{abstract}

\newpage

\section{Introduction}

Some time ago (see \cite{tas1,tas2}), it was possible to prove non
existence of tokamak equilibria with purely poloidal incompressible
flow. Recently, an extension to compressible plasmas appeared in
Ref.\cite{twt} including Hall term and pressure anisotropy. The
proof for the incompressible case given in Refs.\cite{tas1,tas2} was
global while the recent proof \cite{twt} is limited to the
neighbouring of the magnetic axis through a kind of Mercier
expansion.

  This last result motivated the idea to extend the analysis to Vlasov-Maxwell
equations examined near axis. An important ingredient is to write
the Vlasov equation in cylindrical coordinates in a tokamak
geometry, which simplifies the subsequent analysis. We use for that
purpose the calculation done in an old ICTP report \cite{st} where
the Vlasov equation is written in arbitrary orthogonal coordinates.

In Section 2 the expression of the Vlasov equation is obtained in
toroidal geometry. In Section 3 the ODEs of the characteristics are
derived while Section 4 is devoted to "straight tokamaks" and
Section 5 to discussion and conclusions.

\section{Vlasov equation in orthogonal coordinates}

As explained in Ref.\cite{st} we consider a general system of
orthogonal coordinates $x^{1}$, $x^{2}$, $x^{3}$ with the metric
$ds^{2} = g_{11}(dx^{1})^{2} +g_{22} (dx^{2})^{2} +
g_{33}(dx^{3})^{2}$ and unit vectors ${\bf e}_{i} = \frac{\nabla
x^{i}}{|\nabla x^{i}|}$ where i goes from 1 to 3. The velocity
vector of a "microscopic" fluid element is then projected on the
unit vectors ${\bf e}_{i}$ as

\begin{equation}
{\bf v} =  v^{i} {\bf e}_{i},
\end{equation}
where the components $v^{i}$ are independent upon space variables.
The total derivative of ${\bf v}$ is

\begin{equation}
\frac{\partial {\bf v}}{\partial t} + {\bf v}\cdot\nabla {\bf v} =
{\bf E} + {\bf v}\times {\bf B},
\end{equation}
where ${\bf E}$ and ${\bf B}$ are the electric and magnetic fields
consistent with Maxwell equations and the charge to mass ratio
$\frac{e}{m}$ is set to one. Projecting Eq.(2) on the unit vectors
we obtain

\begin{equation}
\frac{dv^{i}}{dt} = {\bf e}_{i}\cdot ({\bf E} + {\bf v}\times {\bf
B}) + {\bf e}_{i}\cdot {\bf v}\times \nabla \times {\bf v}.
\end{equation}
Finally, the Vlasov equation in orthogonal coordinates is given by

\begin{equation}
\frac{\partial f}{\partial t} + {\bf v}\cdot \nabla f + {\bf
e}_{i}\cdot ({\bf E} + {\bf v}\times {\bf B})\frac{\partial
f}{\partial v^{i}} + ({\bf e}_{i}\cdot {\bf v}\times \nabla \times
{\bf v})\frac{\partial f}{\partial v^{i}} = 0,
\end{equation}
where $f$ is a function of the $x^{i}$, $v^{i}$ and time while ${\bf
v}$ is given by Eq.(1). For more details see Ref. \cite{st}. $f$
stays here for the ion distribution while the distribution function
for the electrons is governed by an equation similar to Eq.(4).

Let us now specialize on cylindrical coordinates $x^{1} = r$, $x^{2}
= \phi$, $x^{3} = z$. Then $\nabla \times {\bf e}_{i} = 0$ for $i =
1$ and $3$ and $\nabla \times {\bf e}_{2} = {\bf e}_{1}\times \nabla
\phi$. If we replace the indices 1, 2, 3 by $r, \phi , z$ we have
$\nabla \times {\bf v} = v^{\phi} {\bf e}_{r}\times \nabla \phi$ and

\begin{equation}
{\bf v}\times \nabla \times {\bf v} = \frac{v^{r} v^{\phi}{\bf
e}_{\phi}}{r} - \frac{(v^{\phi})^{2}{\bf e}_{r}}{r}.
\end{equation}
So the last term of Eq.(4) becomes
$[\frac{(v^{\phi})^{2}}{r}\frac{\partial f}{\partial v^{r}} -
\frac{v^{r}v^{\phi}}{ r}\frac{\partial f}{\partial v^{\phi}}]$.
Setting ${\bf B} = {\bf e}_{\phi}\frac{I}{r}$ near axis and
$\frac{\partial f}{\partial t} = 0$ for steady state, Eq.(4) reads

\begin{equation}
{\bf v}\cdot\nabla f + ({\bf e}_{i}\cdot\nabla \Phi)\frac{\partial
f}{\partial v^{i}} -\frac{[v^{z}I -
(v^{\phi})^{2}]}{r}\frac{\partial f}{\partial v^{r}} +
\frac{v^{r}I}{r}\frac{\partial f}{\partial v^{z}} -
\frac{v^{r}v^{\phi}}{r}\frac{\partial f}{\partial v^{\phi}} = 0.
\end{equation}
Assuming $\nabla f = \nabla\Phi = 0$ on axis the final equation to
solve is

\begin{equation}
-[v^{z}I - (v^{\phi})^{2}]\frac{\partial f}{\partial v^{r}} -
v^{r}v^{\phi}\frac{\partial f}{\partial v^{\phi}} +
v^{r}I\frac{\partial f}{\partial v^{z}} = 0.
\end{equation}

\section{ODEs for characteristics}

Let us start with the simpler case $I = 0$, then the characteristics
of Eq.(7) are given by the solution of

\begin{equation}
-\frac{dv^{r}}{(v^{\phi})^{2}} = \frac{dv^{\phi}}{v^{r}v^{\phi}},
\end{equation}
whose solution is $(v^{r})^{2} + (v^{\phi})^{2} = C$. Since $f =
f(C, v^{z}) = f[((v^{r})^{2} + (v^{\phi})^{2}), v^{z}]$ on axis we
obtain for the toroidal flow

\begin{equation}
\int v^{\phi} f d^{3}{\bf v} = 0,
\end{equation}
which means zero toroidal flow on axis.

For $I \neq 0$ the charasteristics are given by

\begin{equation}
-\frac{dv^{r}}{v^{z}I - (v^{\phi})^{2}} =
-\frac{dv^{\phi}}{v^{r}v^{\phi}} = \frac{dv^{z}}{v^{r}I}.
\end{equation}
The last equality delivers $C_{1} = v^{z} + I\ln |v^{\phi}|$, the
second characteristic being the particle energy $C_{2} = (v^{r})^{2}
+ (v^{\phi})^{2} + (v^{z})^{2}$. $C_{1}$ is "antisymmetric" in
$v^{z}$ but symmetric in $v^{\phi}$, which leads to

\begin{equation}
\int v^{\phi} f(C_1, C_2) d^{3}{\bf v} = 0, \int v^{z} f(C_1, C_2)
d^{3}{\bf v} \neq 0.
\end{equation}
It means that the $\phi$-flow is zero while the  unphysical $z$-flow
is finite. This is obviously not acceptable.

\section{"Straight" tokamaks}

The straight tokamaks do have magnetohydrodynamic solutions with
purely poloidal flow as known from previous work \cite{tp}. For the
purpose of a microscopic theory the appropriate coordinate system is
the cartesian one $x^{1} = x$, $x^2 = y$, $x^{3} = z$ so that the
toroidal angular coordinate is replaced by $y$ and the toroidal
field $I$ by $B^{y}$. Since $\nabla\times{\bf e}_{i}$ vanishes for
all $i$, the term ${\bf v}\times\nabla\times{\bf v}$ in Eq.(4)
disappears.

For the steady state with finite $B^{y}$, Eq.(7) is replaced by

\begin{equation}
-v^z\frac{\partial f}{\partial v^x} + v^x\frac{\partial f}{\partial
v^z} = 0,
\end{equation}
whose characteristic is given by
\begin{equation}
-\frac{dv^x}{v^z} = \frac{dv^z}{v^x}.
\end{equation}
The solution of Eq.(13) is $C = (v^x)^{2} +(v^z)^{2}$, which leads
to $f = f((v^x)^{2} +(v^z)^{2}, v^y)$. Purely poloidal flows are
possible, which is consistent with Ref.\cite{tp}.

\newpage

\section{Discussion and Conclusions}

The result of section 3 obliges us to change the assumptions leading
from Eq.(6) to Eq.(7) i.e. $\nabla f \neq 0$ instead of zero on the
magnetic axis. The special canonical $\phi$-momentum solution is of
that kind, and leads naturally to toroidal flows but no poloidal
flows. However, a comprehensive discussion of the problem cannot be
done since the complete set of characteristics of Eq.(6) is not
known.

Finally, though we know from section 3 that $f$ must be a function
of $C_{1}$ and $C_{2}$, we could, in addition, choose $f$ to have
different values for different signs of, for instance, $v^{\phi}$. A
known example of that kind of solutions is the case of BGK waves
\cite{bgk}, in which the "free particles" have different
distributions for different signs of their velocities. See also
Ref.\cite{tas3} for a quasi-neutral treatment. Though toroidal flows
can then be constructed, physical constraints like isotropy of the
pressure tensor or constraints on other moments or geometrical
symmetries and, ultimately, collisions could exclude such solutions.
Again we are led to look for the general solution of Eq.(6) with
$\nabla f \neq 0$ on axis in order to discuss the structure of the
macroscopic flows. Unfortunately, as mentioned before, the answer to
this problem is quite uncertain.

\begin{center}
{\large\bf Acknowledgements}
\end{center}

The authors would like to thank Prof. Harold Weitzner for useful
discussions.

Part of this work was conducted during a visit of the author G.N.T.
to  the Max-Planck-Institut f\"{u}r Plasmaphysik, Garching. The
hospitality of that Institute is greatly appreciated.

The present work was performed under the Contract of Association ERB
5005 CT 99 0100 between the European Atomic Energy Community and the
Hellenic Republic. The views and opinions expressed herein do not
necessarily reflect those of the European Commission.

\end{document}